\documentclass[]{aastex631}

\usepackage{xcolor}
\usepackage{colortbl}
\usepackage{color}
\graphicspath{{./}{figures/}}
\usepackage{hyperref}
\usepackage{float}

\begin{document}

\title{HESS J1809-193: Gamma-Ray Emission by Cosmic Rays from Past Explosion}

\author[0009-0004-1305-9578]{Sovan Boxi}

\email{sovanboxi@rrimail.rri.res.in}
\author[0000-0002-1188-7503]{Nayantara Gupta}
\affiliation{Raman Research Institute \\
C. V. Raman Avenue, 5th Cross Road, Sadashivanagar, Bengaluru, Karnataka 560080, India}
\email{nayan@rri.res.in}

\begin{abstract}
The very high energy gamma-ray source HESS J1809-193 has been detected by the LHAASO and HAWC observatory beyond 100 TeV energy. It is an interesting candidate for exploring the underlying mechanisms of gamma-ray production due to the presence of supernova remnants, pulsar and molecular clouds close to it. We have considered the injection of the energetic cosmic rays from a past explosion, whose reminiscent may be SNR G011.0-00.0, which is located within the extended gamma-ray source HESS J1809-193. We explain the multi-wavelength data from the region of HESS J1809-193 with synchrotron, inverse Compton, bremsstrahlung emission of  cosmic ray electrons and secondary gamma-ray production in interactions of cosmic ray protons with the cold protons in the local molecular clouds within a time dependent framework including the diffusion loss of cosmic rays. The observational data has been modelled with the secondary photons produced by the time evolved cosmic ray spectrum assuming the age of the explosion is 4500 years.

\end{abstract}


\section{Introduction} \label{sec:intro}
The ground based $\gamma$-ray detectors like H.E.S.S. (High Energy Sterescopic System), MAGIC (Major Atmospheric Gamma Imaging Cherenkov), Tibet, HAWC (High Altitude Water Cherenkov) and LHAASO (Large High Altitude Air Shower Observatory) and the space based $\gamma$-ray detectors like {\it Fermi}-LAT and {\it AGILE} are exploring the $\gamma$-ray sky in different energy windows and revealing the very energetic cosmic accelerators in our Galaxy and beyond. They compliment each other in providing us the $\gamma$-ray data covering a few tens of MeV to a thousand PeV energy, and along with the observations at lower frequencies (radio, infrared, optical, ultraviolet and X-ray), it is possible to build up the spectral energy distributions (SEDs) to probe the nature of these sources. H.E.S.S. is an array of high-energy imaging atmospheric Cherenkov telescopes operating in Khomas Highland in Namibia. The galactic plane survey by HESS collaboration has revealed a large population of gamma-ray sources in the very high-energy regime \citep{hesscollab2018ana}. Many of them are either pulsar wind nebulae (PWN) or supernova remnants (SNRs) or composite systems. The extended $\gamma$-ray sources are particularly interesting in this field because it is challenging to identify the different types of sources and different types of radiative mechanisms responsible for the emission in different energy regimes.
\par
The discovery of HESS J1809-193 was reported for the first time in \citet{aharonian2007}. They discussed that the source HESS J1809-193 is located about 0.2$^\circ$ east of the pulsar PSR 1809-1917, which can power it with an efficiency of only $1.2\%$, assuming the $\gamma$-ray source is entirely powered by this pulsar. The possibility that it is powered by a supernova remnant G011.0-00.0 \citep{Bamba_2003} in the X-ray frequency range was also mentioned in this paper. Subsequently, the origin of its emission has been debated in various works due to the presence of SNRs, molecular clouds and pulsar close to the emission region. This source has been observed above 56 TeV and possibly above 100 TeV by HAWC \citep{hawcabeysekara20,hawc_prelim}. Very recently, LHAASO has detected this source with a significance greater than 4-$\sigma$ at an energy exceeding 100 TeV. In the first LHAASO catalogue, this source  has been repoted as an Ultra-High Energy source, designated as 1LHAASO J1908+0615u \citep{cao2023first}. It was also suggested earlier that the pulsar PSR J$1809-1917$ powers the X-Ray PWN of extension $3^{'}$ \citep{kargaltsev2007x, 10.1093/pasj/62.1.179, Klingler_2018, klingler2020chandra}.
\par
\citet{castelletti} discovered a system of molecular clouds at the edge of the shock front of the supernova remnant SNR G011.0-00.0 associated with HESS J1809-193. They suggested that the most likely origin of the very high energy gamma-ray emission is collison of ions accerelated by the SNR in the molecular clouds.
\par
\citet{Araya:2018zsf} modelled the extended GeV emission of HESS J1809-193 with steady state hadronic and leptonic emissions without including the diffusion loss of the cosmic rays. The author included the contributions from the three supernova remnants SNR G011.0-00.0, SNR G11.1+0.1 and SNR G11.4-0.1 to explain the GeV $\gamma$-ray data considering the interactions of cosmic ray protons with the local molecular clouds in the hadronic model and the bremsstrahlung emission in the leptonic model, however the very high energy gamma-ray data was explained in both cases with the inverse Compton emission of the relativistic electrons. 
\par

Recently, \citet{tevhalohess2023} did a detailed observational and theoretical analysis and identified the two components, component A and B of this source. Component A is more extended in space and has a higher $\gamma$-ray flux than component B. They did a time dependent leptonic modelling of HESS J1809-193 including the effect of diffusion loss of the electron and positron pairs injected by the pulsar PSR 1809-1917. The pulsar halo model was used considering two populations of leptons having different ages to explain the multi-wavelength SED. The possibility of hadronic origin of $\gamma$-rays is also discussed in this paper within a steady-state scenario.
\par
\citet{Li_2023} used {\it Chandra} X-ray data to model the SED with leptonic model. They found that a magnetic field of $21\mu$G is needed to explain the extended X-ray halo observed from HESS J1809-193. Due to the high synchrotron emission, inverse Compton emission of the pairs is suppressed; hence, it is hard to explain the very high-energy $\gamma$-ray data with the leptonic model.
\par
We have done a time dependent modeling including the diffusion loss of cosmic rays injected at the site of the source from a past explosion. The secondary photons generated in radiative losses of the cosmic ray electrons and hadronic interactions of cosmic ray protons are used to model the multi-wavelength spectrum of HESS J1809-193. It is shown here that the very high energy gamma-ray components A and B identified in HESS data can be explained in our model by hadronic and leptonic interactions respectively.

\section{HESS J1809-193}
\begin{figure}[h]
    \centering
    \includegraphics[width=0.8\linewidth]{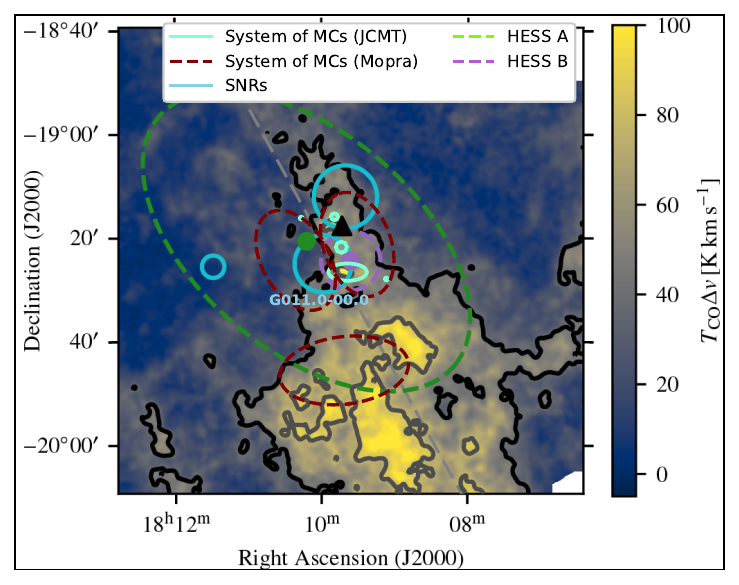}  
    \caption{Positions of SNRs (skyblue circles), FUGIN Map \citep{10.1093/pasj/psx061}, pulsar PSR J1809-1917 (black triangle), HESS A (green dot and dashed contour), HESS B (purple dot and dashed contour) and their 1-$\sigma$ contours from Fig. D1. of \citet{tevhalohess2023}. We have superposed the positions of molecular clouds from JCMT \citep{castelletti} and Mopra \citep{voisin_etal._2019} observations on the same plot. Mopra observation shows three extended molecular clouds with maroon dashed contours.  JCMT observation of MCs due to CO (J$=$3-2) tracer shown in aquamarine contour.}
    \label{fig:mcmorph}
\end{figure}

HESS J$1809\text{-}193$ is an extended TeV gamma ray source located at RA(J2000) = $18^h\, 10^m\, 31^s\ \pm 12^s\ $, Dec(J2000) = -19$^{o}\, 18^{'} \pm \, 2^{'}$ \citep{aharonian2007}. As discussed above, \citet{tevhalohess2023} has modelled this source with two components: the extended asymmetric component as HESS A with 1-$\sigma$ radius of extension of the major and minor axis 
$0.613^{\circ}$, $0.351^{\circ}$ respectively and the symmetric compact component as HESS B with 1-$\sigma$ radius of extension $0.0953^{\circ}$ (see Fig.\ref{fig:mcmorph}). 
Pinpointing the exact counterpart for this source poses a considerable challenge owing to the presence of several potential associations juxtaposed within the depicted region in Fig.\ref{fig:mcmorph}. For instance, radio observation at 330 and 1456 MHz reveals that the region harbours at least two SNRs within the extension of the source, notably G011.1+00.1 and G011.0-00.0 
\citep{green2004galactic, Brogan_2006, castelletti}. The specific distance of G011.1+ 00.1 remains unknown as of now. Estimated distance of G011.0-00.0 \citep{Bamba_2003} is $2.6$ kpc; $2.4$$ \pm$ $0.7$ kpc \citep{Shan_2018}. Unfortunately, the age for SNR G011.0-00.0 is still unresolved \citep{Araya:2018zsf}. The presence of the energetic pulsars PSR J1811-1925 ($\dot{E} = 6.4 \times 10^{36}$ erg s$^{-1}$, $d \sim 5$ kpc) and PSR J$1809\text{-}1917$ ($\dot{E} = 1.8 \times 10^{36}$ erg s$^{-1}$, $d \sim 3.3$ kpc) \citep{manchester2005australia} adds more complexity to the picture. A handful of molecular clouds within the line of sight of the emission region has been independently identified by \citet{castelletti} and \citet{voisin_etal._2019}. We have shown the SNRs and the molecular clouds (MCs) in Fig.\ref{fig:mcmorph} along with the pulsar PSR J1809-1917. $^{12}$CO($3-2$) observations carried out by \citet{castelletti} using the James Clerk Maxwell Telescope (JCMT, Mauna kea, Hawaii) reveal a substantial overlap between the primary emission region of the HESS source and molecular cloud system. Deep radio observation around the source using the Karl G. Jansky Very Large Array, JVLA by the same group also confirms  a strong spatial correlation between the shock front of the SNR G011.0-00.0 and molecular cloud system. Using HI 21 cm absorption technique and $^{12}$CO emission they have estimated the distance of the molecular clouds and by HI absorption the distance of SNR G011.0-00.0, and subsequently established a physical connection between them. The average distance of this cloud system as well as the associated SNR as quoted there is $\approx 3$kpc. The predicted mass of each individual cloud varies from $7 \times 10^{2}-1.3 \times 10^{3} M_{\odot}$ for velocity 21 km/sec  \citep[Table 2]{castelletti}. Likewise, the intensity map CS($1-0$) using Mopra Telescope \citep{voisin_etal._2019} covering various  LSR (local standard of rest) velocities of molecular material unveils the existence of  clumps of molecular clouds stretched over the vicinity of SNRs and the emission region.  Depending on the tracer selected, mass of individual cloud differs, ranging from $10^3- 2.3 \times 10^5 M_{\odot}$. In similar fashion, molecular hydrogen density varies from $150$ cm$^{-3}$ to $4.4 \times 10^4$ cm$^{-3}$ (Table $D.1$ \citet{voisin_etal._2019}). We have presented an intensity map superposing the JCMT \citep{castelletti} and Mopra \citep{voisin_etal._2019} observations over FUGIN map \citep{10.1093/pasj/psx061} in Fig.\ref{fig:mcmorph}. 
\par
It is evident from our Fig.\ref{fig:mcmorph} as well as the discussion above that the molecular hydrogen density varies by orders of magnitude over the entire emission region. The spatial variation in matter distribution in the region of HESS J1809-193 suggests the interaction of SNR shock and molecular clouds may have an important role in powering this source. The cosmic ray protons lose energy slower than the cosmic ray electrons, so they can propagate to longer distances before losing energy and may produce extended emission (HESS component A) in very high energy gamma-rays, while cosmic ray electrons can emit very high energy gamma-rays by inverse Component mechanism and produce HESS component B of HESS J1809-193 \citep{tevhalohess2023}.

\section{SED Modelling}
\begin{figure*}[h]
\gridline{\fig{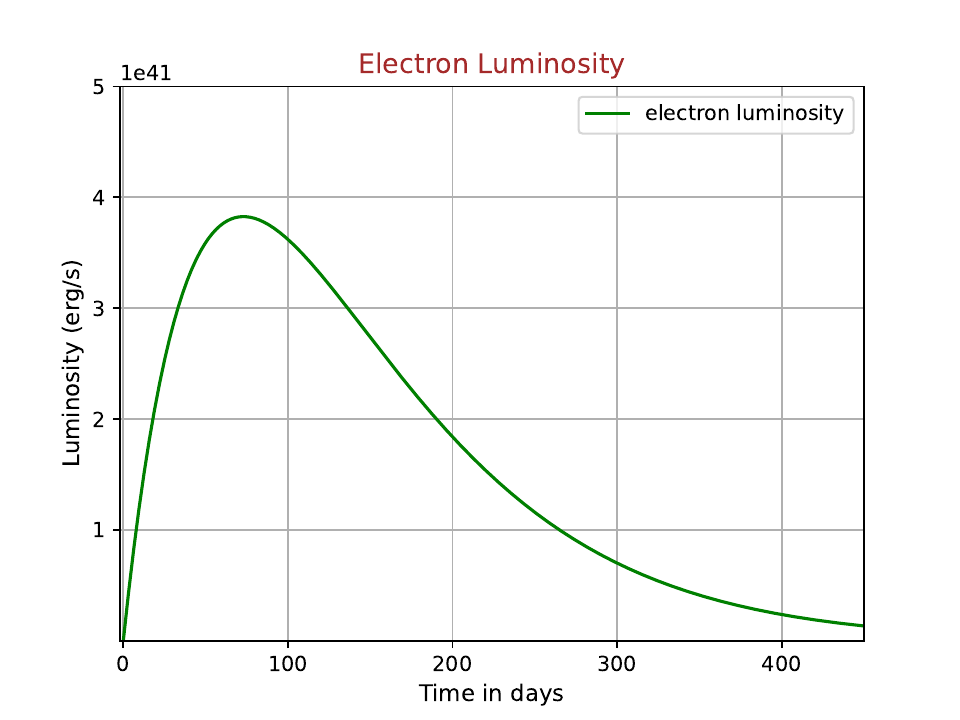}{0.5\textwidth}{(a)}\\
          \fig{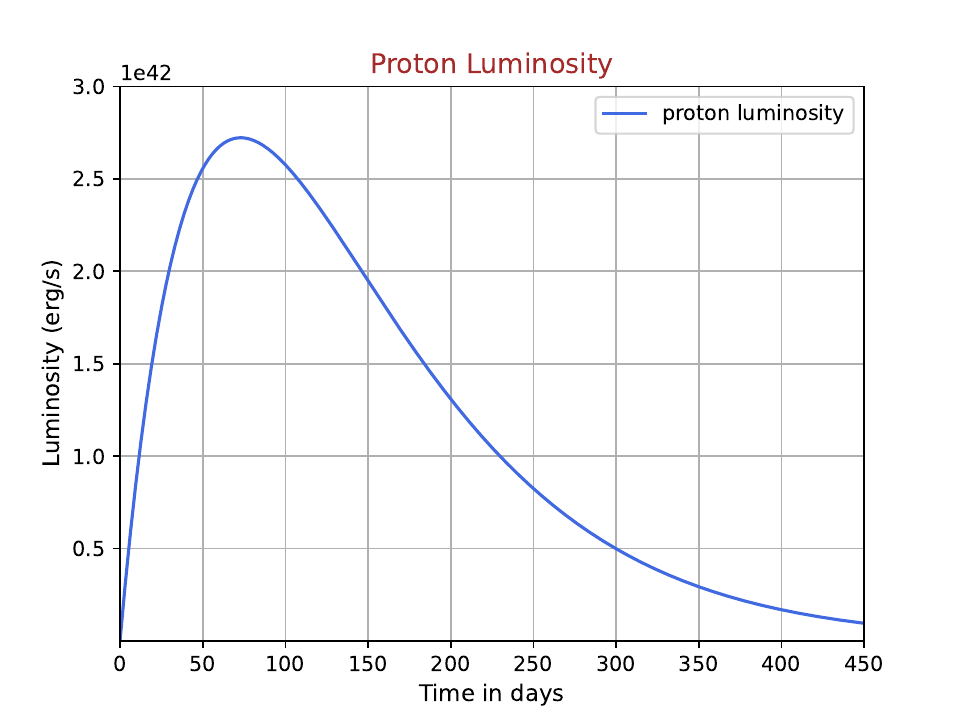}{0.5\textwidth}{(b)}\\}
\caption{ Time evolution of injected luminosity in electrons (a) and protons (b)} 
\label{fig:lcinject}
\end{figure*}

In this section, we discuss the modelling of the observed multi-wavelength spectrum in the region of HESS J1809-193. The cosmic rays are accelerated in supernova shocks by diffusive shock acceleration(DSA) \citep{BLANDFORD19871} mechanism. We have assumed that the supernova explosion at the location of the source lasted for nearly a year and injected very high-energy cosmic ray electrons and protons, which have power law distributions in energy. Subsequently, they lose energy in the ambient magnetic field, radiation field and matter, and some escape from the emission region. The transport equation is given below, used to calculate the time-evolved cosmic ray electron and proton spectra. The electrons are losing energy by synchrotron emission in the local ambient magnetic field and inverse Compton (IC) emission in the local interstellar radiation field (ISRF) and cosmic microwave background (CMB) radiation. The electrons also lose energy by bremsstrahlung emission in the molecular clouds. The protons are losing energy mainly in proton-proton interactions with the molecular clouds. The densities of the molecular clouds vary by several orders of magnitude over the emission region, hence we have used an average value to fit the observed gamma-ray data. We have used the GAMERA code \citep{2022ascl.soft03007H} to calculate the time evolved cosmic ray particle spectra, including all the energy loss processes and the escape of cosmic rays, and subsequently the non-thermal emission covering radio to very high energy gamma-ray frequencies from the time evolved particle spectra.
The transport equation used to study the time evolution of the relativistic electrons or protons is given by
\begin{equation} 
\frac{\partial{N(E,t)}}{\partial{t}}=Q(E,t)-\frac{\partial{[b(E,t)N(E,t)]}}{\partial{E}}-\frac{N(E,t)}{t_{diff}}
\end{equation}
 where $N(E,t)$ is the resulting particle spectra at any time t, $Q(E,t)$ is the injection spectra, $b=b(E,t)$ represents the energy loss of particles. The exact forms of the time-dependent injected luminosity in electrons and protons are shown separately in Fig.\ref{fig:lcinject}(a) and Fig.\ref{fig:lcinject}(b), respectively.
  In GAMERA framework, the complete Klein-Nishina cross-section for IC scattering \citep{RevModPhys.42.237} is utilized to compute the photon flux resulting from the relativistic electrons. For calculating the IC emission from the ISRF target photon field, we have used the distribution of ISRF as given in \citet{10.1093/mnras/stx1282}. The code calculates the bremsstrahlung radiation for both electron-electron and electron-ion interactions. For proton-proton interactions, GAMERA uses semi-analytical parametrisation developed by \citet{Kafexhiu:2014cua}. In our calculation, we have incorporated the GEANT 4.10.0 hadronic interaction model for proton-proton interaction. We have included the energy-dependent diffusion loss of cosmic ray electrons and protons using the following form of diffusion coefficient.
\begin{equation} \label{trans}
D = D_0  \Big(\frac{E}{E_0}\Big)^{\delta}   
\end{equation}
where $D_o=1.1 \times 10^{28}$ cm$^{2}$ s$^{-1}$ is the diffusion coefficient normalised at $E_0 = 40$ TeV and $\delta = 0.58$ following \citep{tevhalohess2023}. They fitted the observed size of component A with the help of the GAMERA library \citet{2022ascl.soft03007H} and the best-fitted model provides the value for $\delta$, $D_{o}$ \citep{tevhalohess2023}, this value of $D_{o}$ is similar to that reported in earlier work \citep{abeysekara2017extended}. We have used these best-fitted values in our calculation.
  The diffusion-loss mechanism of particles (electrons and protons) has been included through the diffusion time-scale $t_{diff}$ in the transport equation. The diffusion timescale $t_{diff}$ can be expressed in terms of the diffusion coefficient $D$ and the size of the region where the cosmic rays are trapped and lose energy, which we assume to be the same as the size of the emission region.
\begin{equation}\label{diff}
    t_{diff}=\frac{L_{emission}^2}{D}.
\end{equation}

\begin{figure*}[ht]
\gridline{\fig{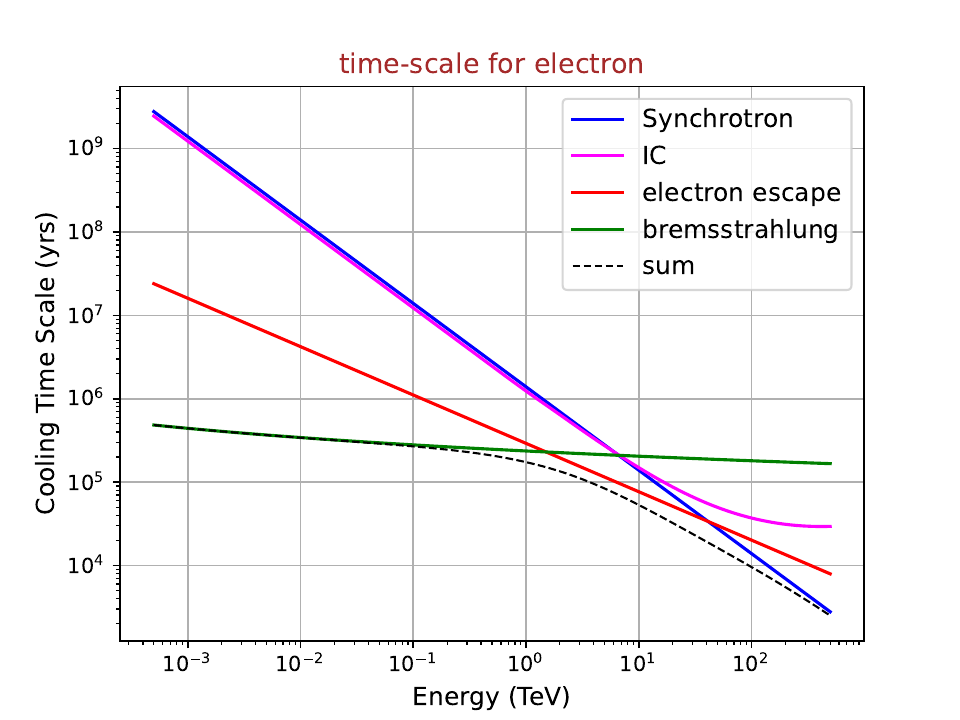}{0.5\textwidth}{(a)}\\ \fig{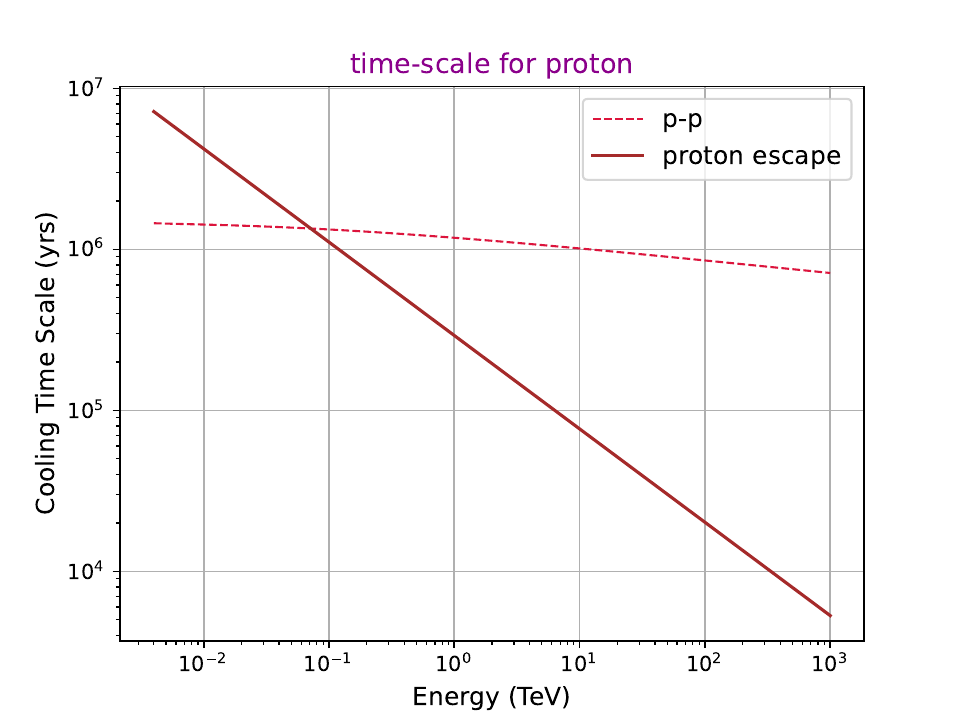}{0.5\textwidth}{(b)}}
\caption{ Energy loss timescales of electrons (a) and protons (b).}
\label{fig:cooling}
\end{figure*}

\begin{figure}[ht]
    \centering
    \includegraphics[width=0.9\linewidth]{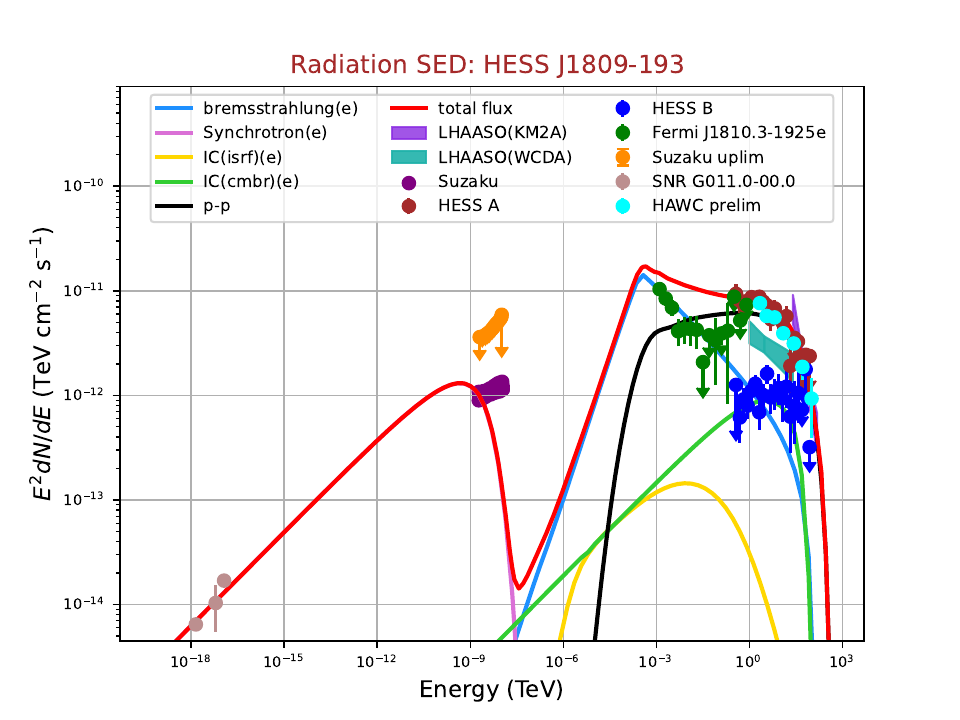}  
    \caption{Multiwavelength SED of HESS J1809-193. Datapoints from different observations: Fermi-LAT in green dot \citep{tevhalohess2023}, two-component HESS data: HESS-A in brown and HESS-B in blue \citep{tevhalohess2023}; HAWC preliminary data in cyan \citep{hawc_prelim}, LHAASO: WCDA data between 1$-$25 TeV in lightseagreen butterfly and KM2A data beyond 25 TeV in purple butterfly \citep{cao2023first}; X-Ray data from Suzaku: butterfly in violet and upper limit in orange  \citep{10.1093/pasj/62.1.179}; Radio data for SNR G011.0-00.0 in rosybrown. \citep{Brogan_2006}}
    \label{fig:sed}
\end{figure}

In the above equation, we assume $L_{emission}$ to be the total extension of the emission region observed by HESS, which is approximately 37 parsecs.   
The escape and cooling time-scales of electrons and protons have been shown in Fig \ref{fig:cooling}(a) and Fig \ref{fig:cooling}(b), respectively, for the regions near HESS J1809-193. At higher energy, synchrotron cooling of electrons dominates over other energy loss processes. The diffusion loss of protons is very important above 100 GeV energy as the diffusion loss time-scale decreases rapidly with increasing energy. The age of the explosion is adjusted along with the other parameters listed in Table 1 to fit the SED shown in Fig.\ref{fig:sed}.

\begin{deluxetable*}{|c|c|}[h]
\tablenum{1}
\tablecaption{Parameters used for modelling HESS J1809-193\label{tab:parameter}}
\tablewidth{0pt}
\tablehead{
\colhead{Parameters} & \colhead{HESS J1809-193} 
}
\startdata
Energy injected in protons & $4.67\times 10^{49}$ erg  \\
Energy injected in electrons &$6.56 \times 10^{48}$ erg \\
Maximum energy of protons injected &$10^{3}$ TeV \\
Minimum energy of protons injected& $4\times 10^{-3}$ TeV  \\
Maximum energy of electrons injected & $5\times 10^{2}$ TeV  \\
Minimum energy of electrons injected &$5\times 10^{-4}$ TeV \\
Spectral index of injected proton spectrum &$-2$ \\
Spectral index of injected electron spectrum &$-2.4$ \\
Number density of particles in molecular cloud & $50$ cm$^{-3}$ \\
Magnetic field in emission region & $3$ $\mu $G   \\
Distance of molecular cloud & 2.6 kpc \\
Age of explosion & $4500$ years  \\
\enddata

\end{deluxetable*}

\section{Results}
 Fig.\ref{fig:mcmorph} strongly suggests association between SNRs and the molecular cloud system. We have used this association in modelling the source HESS J1809-193 within a time-dependent framework. We have assumed that SNR G011.0-00.0, whose age is not known, exploded thousands of years ago and cosmic ray electrons and protons were injected in the region of HESS J1809-193 from this explosion. The cosmic rays have cooled down by interactions with the ambient magnetic field, radiation field and matter distributed near the SNR. The cosmic ray electrons have lost their energy in synchrotron, inverse Compton with ISRF, CMB photons and bremmstrahlung emission and the cosmic ray protons have lost energy in proton-proton interactions with the ambient hydrogen molecules. The synchrotron self Compton (SSC) emission is very low compared to the emission from the other processes but it has been included in our work.
 
 We have also assumed that the injection luminosity was time-varying and the injection of cosmic rays continued for a year. The precise form of the injected luminosity used in this work has been shown in Fig.\ref{fig:lcinject}. 
 We have used the publicly available software GAMERA to calculate the time-evolved cosmic ray spectrum and photon spectrum at present-day after including radiative losses of electrons, hadronic interactions and diffusion loss of cosmic rays. The radiation spectra of electrons and protons vary with the explosion's age and the total energy injected during the explosions. After adding the leptonic and hadronic contributions, the values of these parameters are varied so that the total SEDs fit the observed spectra. In Fig.\ref{fig:sed}, we show the total SEDs and the individual components in our lepto-hadronic model. The values of the parameters used in our model are listed in Table 1.
The following spectral distributions of injected cosmic rays: (a) electrons: $\frac{dN_e}{dE_e}\propto {E_{e}}^{-2.4}$ and (b) protons: $\frac{dN_p}{dE_p}\propto {E_p}^{-2}$ have been adapted to fit the SEDs. The total amount of energy injected into particles throughout the age of the source to explain the observed data at the present day is (a) $W^{p} = 4.67\times 10^{49} $ erg for the cosmic ray protons; and (b) $W^{e} = 6.56 \times 10^{48} $ erg for the cosmic ray electrons; their sum is only a few percent of the kinetic energy released in a canonical supernova explosion $\sim 10^{51}$ erg \citep{Ginzburg_1975}.

 \par
 
 As shown in Fig.\ref{fig:sed} the radio data points from SNR G011.0-00.0 can be well explained by the synchrotron emission from the cosmic ray electrons in our model. The synchrotron emission in the X-ray band passes through the butterfly region of the Suzaku observation, but does not cover the entire region. Fermi LAT data has been fitted using the bremmstrahlung emission of the cosmic ray electrons in molecular clouds and proton-proton interactions. The HESS data points for the B component can be explained with inverse Compton emission of the cosmic ray electrons, and the extended A component with proton-proton interactions. The HAWC data points, HESS data points for the A component and LHAASO KM2A data points overlap with each other and they are well fitted with hadronic interactions in our model. LHAASO WCDA data points near 25 TeV partially overlap with the HESS data points for the A component. At lower energy the LHAASO WCDA data points are in between the data points of B and A components of HESS.
Thus it is possible to fit most of the $\gamma$-ray data points with a single source, however it is also possible that more than one source may contribute to the observed emission as there are multiple sources in the region of HESS J1809-193.
 


\section{Discussion and Conclusion}
HESS J1809-193 is an interesting source to study the multi-wavelength emission mechanisms in the particle acceleration site due to the availability of sufficient observational data covering radio to hundred TeV $\gamma$-ray energy from different observatories. SNRs, pulsar and molecular clouds are present at the location of the source, this association suggests different possible scenarios for the underlying emission mechanisms. Earlier works have considered leptonic \citep{aharonian2007,tevhalohess2023} models to explain the multi-wavelength data with the electrons emitted by the pulsar and also hadronic model has been applied to explain the source HESS J1809-193 within a steady state scenario \citep{Araya:2018zsf}.
 \par
We have suggested that the explosion of SNR G011.0-00.0 may be the origin of HESS J1809-193. We have explained the multi-wavelength data within a time dependent framework after including all the energy loss process and the diffusion loss of cosmic rays. In our model the maximum energy of the injected protons is higher than the maximum energy of the injected electrons because the electrons cool down much faster than the protons and it is much harder to accelerate the electrons to very high energy. Since, protons lose energy slowly and propagate to a longer distance before losing energy, they are expected to give the emission from the extended region at the highest observed energy. It has been discussed earlier that the spatial distribution of the molecular clouds is not reflected in the extended $\gamma$-ray emission observed by HESS \citep{tevhalohess2023}. We note that the energy dependent diffusion time scale of the cosmic ray protons is much shorter than the proton-proton interaction time-scale at very high energy, and due to this reason the distribution of the protons is not uniform in the emission region, as a result the $\gamma$-ray emission may not trace the distribution of molecular clouds. It would be possible to study the morphology and the distribution of $\gamma$-ray emission at the site of HESS J1809-193 in more detail with more observational data in future. There may be multiple sources working together at the site of HESS J1809-193 and the role of the pulsar PSR J1809-1917 to produce a pulsar halo may be important. Our work presents an alternative scenario to explain HESS J1808-193.

\section*{Acknowledgement}
The authors thank the referee for helpful comments.

\software{GAMERA (\url{https://ascl.net/2203.007})}

\bibliography{reference}{}
\bibliographystyle{aasjournal}

\end{document}